\documentstyle[12pt,twocolumn]{article}

\begin{document}

\setlength{\textwidth}{480pt}
\setlength{\textheight}{630pt}
\setlength{\topmargin}{-0.375in}
\setlength{\oddsidemargin}{-0.0833in}
\setlength{\evensidemargin}{-0.0833in}
\setlength{\parindent}{0in}
\setlength{\parskip}{0pt}

\title{
On the Equation $H=mv^2$ and the Fine Structure of the Hydrogen Atom
}

\author{
{\large Ezzat G. Bakhoum}\\
\\
{\normalsize New Jersey Institute of Technology}\\
{\normalsize P.O. Box 305, Marlton, NJ. 08053 USA}\\
{\normalsize Email: bakhoum@modernphysics.org}\\
\\
{\normalsize Copyright \copyright 2002 by Ezzat G. Bakhoum}
}

\date{\normalsize Posted: July 26, 2002 \hspace*{.25in} Updated: December 26, 2002}

\maketitle

{\large\bf Abstract:}\\
\\
The recently introduced\cite{Bakhoum} reconciliation of the theories of special relativity and wave mechanics implies that the mass-energy equivalence principle must be expressed mathematically as $H=mv^2$, where $H$ is the total energy of a particle, $m$ is its relativistic mass, and $v$ is its velocity; not $H=mc^2$ as was widely believed. In this paper, the equation $H=mv^2$ will be used to calculate the energy levels in the spectrum of the hydrogen atom. It is demonstrated that the well-known Sommerfeld-Dirac formula is still obtained, but without the constant term $m_0 c^2$ that was originally present in the formula.

 
\pagebreak

{\large\bf 1. Introduction:}\\
\\
In an earlier work by the author\cite{Bakhoum} it was demonstrated that a total energy equation that will satisfy the Compton-de Broglie wave mechanics as well as the theory of special relativity will be the equation $H=mv^2$ (where $H$ is the total energy of a particle, $m$ is its relativistic mass, and $v$ is its velocity); not the equation $H=mc^2$ as was widely accepted. In this paper, we will seek to calculate the fine structure of the spectrum of the hydrogen atom, using the equation $H=mv^2$.\\
\\
According to the improved relativistic model of the atom advanced by the theories of Sommerfeld\cite{Sommerfeld} and Dirac\cite{Dirac}, the energy levels in the spectrum of the hydrogen atom were found to be in agreement with the following approximate equation

\begin{equation}
W \approx m_0 c^2 - 
\frac{m_0 e^4}{2 \hbar^2 \left(n - k + \sqrt{k^2 - \alpha^2}\right)^2},
\label{SD}
\end{equation}

where $n$ is the principal quantum number, $k$ is the second (or Sommerfeld's) quantum number, and $\alpha$ is Sommerfeld's fine structure constant. As it is well recognized, this equation describes the fine structure accurately, except for the Lamb shift\cite{Schweber},\cite{Kragh}. The constant term $m_0 c^2$ present in the equation, though it was not present in the original theory of Bohr, does not of course affect the observed energy levels since it is canceled when the difference $W_1 - W_2$ between any two levels is taken. In the following sections we will demonstrate, by using the total energy equation $H=mv^2$, that Eq.(\ref{SD}) is still obtained except for the constant term $m_0 c^2$. Even though the analysis is still fully relativistic, the absence of the term $m_0 c^2$ makes the final equation essentially an improved version of the original equation derived by Bohr.\\
\\
We have two alternatives in proceeding with this analysis: we can either use the simple and elegant geometric approach developed by Sommerfeld, or use the Hamiltonian approach developed by Dirac. We will opt for the second choice for one simple reason: Dirac's analysis did depend on Einstein's total energy formula $H = mc^2$. It is therefore very important to examine how the analysis and hence the final result will be shaped when the total energy is taken to be $H=mv^2$ instead. In the analysis, we will of course assume a heavy, spinless nucleus and no radiative corrections. We will also follow the standard practice of assigning the symbol ``$e^2$'' as a shorthand for the quantity $e^2/4\pi\epsilon_0$ that appears in Coulomb's law (where $e$ is the electron's charge).\\
\\
\\
{\large\bf 2. The relationship between the equation $H=mv^2$ and Bohr's model of the atom:}\\
\\
Apart from the relativistic mechanics, the equation $H=mv^2$ has a very strong relationship with the classical mechanics used by Bohr in his model of the atom. It is important to understand this relationship before we proceed with a Hamiltonian-based analysis. In Bohr's model of the atom, the electron's kinetic and potential energies are both equal to zero at an infinite distance from the nucleus. As the electron moves closer to the nucleus, it loses potential energy and hence this quantity becomes negative. This quantity is defined by Coulomb's law and is numerically equal to $- e^2/R$ in the case of the hydrogen atom, where  $R$ is the orbit's radius. There is, however, an alternative way to obtain a mathematical expression that represents the same quantity. Since a free electron traveling with a linear velocity $v$ at an infinite distance from the nucleus must have a total energy equal to $m v^2$, then the principle of conservation of energy implies that a bound electron traveling with the same velocity $v$ must have a potential energy equal to $-mv^2$ (the amount of energy required to free the electron must be equal to the total energy available from the free particle). It is not difficult to see that this simple result is in perfect agreement with the classical mechanics in Bohr's theory: classically, the electron is in equilibrium due to the equality of the two forces

\begin{equation}
\frac{e^2}{R^2} = \frac{m v^2}{R},
\label{11}
\end{equation}

where $e^2/R^2$ is the Coulomb electrostatic force and $mv^2/R$ is the centrifugal force. It is now obvious that the quantity $- e^2/R$, or the potential energy, is numerically equal to the quantity $-mv^2$. Two observations are now immediately apparent: first, given that the potential energy is numerically equal to $-mv^2$, then the total energy of a bound electron must be equal to zero; secondly, it will be obvious that the equation $H = mv^2$, while fully in agreement with the theories of special relativity and wave mechanics, it is also fully distinguishable from the classical theory of Bohr.\\
\\
It is now important to understand how -given that $H=0$- the spectral lines of the atom are obtained. Since exciting an atom means applying a certain amount of external work to the electron, then the ``observable'' energy, that is, the energy that can be transferred to/from the atom, must be equal to the sum of the kinetic and potential energies. This was the assumption that was used by Bohr and later Sommerfeld and Dirac. We will also preserve the same principle here. We will refer to that observable energy as $W$. It was demonstrated in the earlier paper by the author\cite{Bakhoum} that the kinetic energy of a particle is given by the following relativistic expression:   

\begin{equation}
E_k = \Delta m c^2 = m v^2 - m_0 c^2 (1 - \sqrt{1 - v^2/c^2}). 
\label{12}
\end{equation}

The observable energy $W$, therefore, will be given by

\begin{eqnarray}
W & = & \mbox{K.Energy + P.Energy}\nonumber\\
  & = & \left[ m v^2 - m_0 c^2 (1 - \sqrt{1 - v^2/c^2}) \right] -           mv^2\nonumber\\
  & = & - m_0 c^2 (1 - \sqrt{1 - v^2/c^2}).
\label{13}
\end{eqnarray}

The fact that $W$ is a negative quantity is, of course, not new. What is new here is that $W$ is numerically equal to the electron's mass-energy quantity! This result has an important physical meaning: for a bound electron, what is classically referred to as the ``sum of kinetic and potential energies'' is, from a relativistic point of view, a quantity that actually represents the mass-energy of the particle. To understand it differently, if we consider the atom to be a ``system'' that is capable of storing energy, then any excitation energy applied to that system will be stored in the form of additional mass-energy of the moving particle, that is, the electron. It is important to observe that this conclusion correlates strongly with the principles of wave mechanics, since classical quantities such as ``kinetic energy'' and ``potential energy'' have no equivalent in wave mechanics. Now, mathematically, this result means that while the total energy of the electron -as a standalone particle- is zero, the energy stored in the {\em system}, that is, the atom, is different from zero.\\ 
\\
The only unknown in Eq.(\ref{13}) is of course the velocity $v$. We shall now proceed to calculate $v$, using the Hamiltonian technique developed by Dirac. We will then be able to write a formula for the observable energy $W$ in terms of quantum numbers.\\
\\
\\
{\large\bf 3. The energy level $W$ based on the Hamiltonian analysis:}\\
\\
We will begin by considering the Hamiltonian of a free particle, obtained in the earlier paper\cite{Bakhoum}:

\begin{equation}
\vec{H} = \pm v \; \sum_r \vec{p_r} \: [\beta_r],
\label{21}
\end{equation}

where $\vec{p_r}$ is a vectorized one-dimensional momentum component, $[\beta_r]$ is a Dirac matrix, and $r=1,2,3,4$. For a bound electron, we just add the potential energy $-e^2/R$ to the Hamiltonian. But since $-e^2/R = -mv^2$ then $H=0$ as we indicated, and hence the eigenvalues of $\vec{H}$ are also zeros. We can therefore write the following scalar equation

\begin{equation}
\left( \pm v \; \sum_r \vec{p_r} \: [\beta_r] \right)^2 -
\left( - \frac{e^2}{R} \right)^2 = 0
\label{21b}
\end{equation}

In Dirac's analysis\cite{Dirac}, he had demonstrated that the sum $\pm \sum_r \vec{p_r} \: [\beta_r]$ can be represented in the radial direction (that is, the direction extending from the nucleus to the electron) by the following linear operator:

\begin{equation}
\pm \; \sum_r \vec{p_r} \: [\beta_r] = 
\pm \hbar \: \left( \frac{\partial}{\partial R} \pm \frac{k}{R} \right),
\label{22}
\end{equation} 

where $k$ is a second quantum number (to be differentiated from the main quantum number $n$, which will be used later. Here, $k$ is actually Sommerfeld's quantum number, since in Dirac's original notation he used $1\pm j$ instead of $\pm k$). In addition, it is easy to demonstrate, as Dirac has shown, that any two scalars $A$ and $B$ that satisfy the relationship $A^2 - B^2 = 0$ must also satisfy the two equations

\begin{eqnarray}
A \: \psi_a + B \: \psi_b & = & 0 \nonumber\\
A \: \psi_b + B \: \psi_a & = & 0,
\label{22b}
\end{eqnarray}

where $\psi_a$ and $\psi_b$ is another set of two different scalars. In view of these definitions, Eq.(\ref{21b}) can now be written in the form of two equations:

\begin{eqnarray}
\pm \hbar v \: \left( \frac{\partial}{\partial R} \pm \frac{k}{R} \right) \: \psi_a - \frac{e^2}{R} \: \psi_b & = & 0, \nonumber\\
\pm \hbar v \: \left( \frac{\partial}{\partial R} \pm \frac{k}{R} \right) \: \psi_b - \frac{e^2}{R} \: \psi_a & = & 0
\label{23}
\end{eqnarray}

Here, $\psi_a$ and $\psi_b$ represent two different radial wave functions of the electron. We now define the following new quantity, let

\begin{equation}
\alpha = \frac{e^2}{\hbar v}.
\label{23b}
\end{equation}

(Note that $\alpha$ represents a special case of Sommerfeld's fine-structure constant, $e^2/\hbar c$, which will emerge later). Eqs.(\ref{23}) are now rewritten as

\begin{eqnarray}
\left( \frac{\partial}{\partial R} \pm \frac{k}{R} \right) \: \psi_a \pm \frac{\alpha}{R} \: \psi_b & = & 0, \nonumber\\
\left( \frac{\partial}{\partial R} \pm \frac{k}{R} \right) \: \psi_b \pm \frac{\alpha}{R} \: \psi_a & = & 0
\label{24}
\end{eqnarray}

We next consider the structure of the radial wave functions $\psi_a$ and $\psi_b$. It was demonstrated\cite{Dirac},\cite{French} that $\psi$, to be convergent for large $R$ as well as for small $R$, it must be of the form

\begin{equation}
\psi = \left( \sum_s C_s R^s \right) \: e^{- \gamma R}
\label{24b}
\end{equation}

where $\gamma$ is a constant, $C_s$ are constant coefficients, and where $s$ represents increasing powers of $R$. We will follow Dirac's representation by using the following convenient form for $\psi_a$, $\psi_b$:

\begin{equation}
\psi_a = e^{- \gamma R} \: f, \qquad \psi_b = e^{- \gamma R} \: g,
\label{24c}
\end{equation}

where $f= \sum_s C_s R^s$, $g= \sum_s C_s^\prime R^s$ are two different power series. Eqs.(\ref{24}) will now become

\begin{eqnarray}
\left( \frac{\partial}{\partial R} \pm \frac{k}{R} - \gamma \right) \: f
\pm \frac{\alpha}{R} \: g & = & 0 \nonumber\\
\left( \frac{\partial}{\partial R} \pm \frac{k}{R} - \gamma \right) \: g
\pm \frac{\alpha}{R} \: f & = & 0,
\label{25}
\end{eqnarray}

or, by substituting with the equivalent expressions for $f$ and $g$,

\begin{eqnarray}
\lefteqn{\sum_s s C_s R^{s-1} \pm k \sum_s C_s R^{s-1} - } \qquad \nonumber\\
 & & \gamma \sum_s C_s R^s \pm \alpha \sum_s C_s^\prime R^{s-1} = 0,\nonumber\\
\lefteqn{\sum_s s C_s^\prime R^{s-1} \pm k \sum_s C_s^\prime R^{s-1} - } \qquad \nonumber\\
 & & \gamma \sum_s C_s^\prime R^s \pm \alpha \sum_s C_s R^{s-1} = 0 \nonumber\\
                                           &   & 
\label{26}
\end{eqnarray}

We will assume, along with Dirac, that the first term in the wave function power series starts with the power $s_0$, so that the sum of the coefficients of $R^{s_0-1}$ in the above equations must be equal to zero. By selecting the coefficients of $R^{s_0-1}$ we therefore get the following two equations:

\begin{eqnarray}
s_0 C_{s_0} \pm k C_{s_0} \pm \alpha C^\prime_{s_0} & = & 0 \nonumber\\
s_0 C^\prime_{s_0} \pm k C^\prime_{s_0} \pm \alpha C_{s_0} & = & 0,
\label{27}
\end{eqnarray}
  
or,

\begin{eqnarray}
\frac{C_{s_0}}{C^\prime_{s_0}} \: (s_0 \pm k) & = & \pm \alpha, \nonumber\\
\frac{C^\prime_{s_0}}{C_{s_0}} \: (s_0 \pm k) & = & \pm \alpha
\label{28}
\end{eqnarray}

If we now multiply the respective terms on each side of these two equations, while allowing for a different choice of the signs of $\alpha$ and $k$ in each equation, we get 

\begin{equation}
- \alpha^2 = (s_0 + k) (s_0 - k) = s_0^2 - k^2,
\label{29}
\end{equation}

hence,

\begin{equation}
s_0 = \sqrt{k^2 - \alpha^2}
\label{29b}
\end{equation}

The relationship between $\alpha$ and the power $s$ has an important physical meaning. $\alpha$ is a function of the velocity of the electron, while the equations (\ref{26}) represent spherical harmonics in which increasing powers (that is, increasing $s$) correspond to different electron positions (i.e., increasing $s$ correspond to increasing $n$, or the principal quantum number). We must therefore put $\alpha_s$ instead of $\alpha$ in those two equations, since $\alpha$ will be a variable in those spherical harmonics, given of course the fact that the velocity will be a function of the position, or the principal quantum number. For the power $s_0$, which is the first term in the series, we shall designate the symbol $\alpha_0$ as the value of $\alpha$ that corresponds to that term. Eq.(\ref{29b}) is therefore written as

\begin{equation}
s_0 = \sqrt{k^2 - \alpha_0^2}
\label{29c}
\end{equation}

We next turn to the following question: what is the numerical value of $\alpha_0$ ? The answer is in fact quite simple. Since $\alpha = e^2 / \hbar v$, and since the power $s_0$ corresponds to the very first term in the spherical harmonic series, a term which in turn corresponds to a position of the electron that is the lowest possible theoretically, then $v$ must have the maximum possible theoretical value, or $c$. $\alpha_0$ will then be given by

\begin{equation}
\alpha_0 = \frac{e^2}{\hbar c}.
\label{210}
\end{equation}

We should now recognize that $\alpha_0$ is actually Sommerfeld's fine-structure constant. The next step is to determine the value of $\alpha_s$ when the power $s$ is at a maximum, which corresponds to the last term in the series and hence the actual position of the electron. In Eqs.(\ref{26}), the coefficients $C_s$ and $C^\prime_s$ must both tend to zero if the series is to converge for a certain value of $s$. If we now select the powers of $R^{s-1}$ in the two equations and let $C_s \approx C^\prime_s \approx 0$, we get the following single equation

\begin{equation}
\alpha_s = \pm (s \pm k)
\label{211}
\end{equation}

We now have the following restrictions on the three quantities $\alpha_s$, $s$ and $k$: all three are positive quantities; $k \leq s$; $s$ must be larger than the principal quantum number $n$; and finally $\alpha_s$ cannot be much larger than $n$ or otherwise the velocity $v$ will deviate strongly from the prediction of the classical theory. These restrictions immediately lead to the conclusion that the only possible solution for Eq.(\ref{211}) is $\alpha_s = (s-k)$. We finally let $s = n + s_0$, as Dirac had proposed. The quantity $\alpha_s$ will therefore be given by

\begin{equation}
\alpha_s = s - k = (n + \sqrt{k^2 - \alpha_0^2}) - k
\label{212}
\end{equation}

The velocity of the electron will therefore be

\begin{equation}
v = \frac{e^2}{\hbar \alpha_s} = \frac{e^2}{\hbar \: (n - k + \sqrt{k^2 - \alpha_0^2})}
\label{213}
\end{equation}

From Eq.(\ref{13}), the observable energy $W$ will now be given by

\begin{eqnarray}
\lefteqn{W =  -m_0 c^2 \; (1- } \nonumber\\
& & \left[
1- \frac{e^4}{\hbar^2 c^2 \left(n - k + \sqrt{k^2 - \alpha_0^2}\right)^2}
\right]^{1/2}
\; ) \nonumber\\
 & &
\label{214}
\end{eqnarray}

Since the ratio $v^2/c^2$ is much smaller than unity, the equation can be written approximately as

\begin{eqnarray}
W & \approx & - \: m_0 c^2 \left[ 
\frac{e^4}{2 \hbar^2 c^2 \left(n - k + \sqrt{k^2 - \alpha_0^2}\right)^2}\right]
\nonumber\\
  & \approx & 
- \: \frac{m_0 e^4}{2 \hbar^2 \left(n - k + \sqrt{k^2 - \alpha_0^2}\right)^2} \nonumber\\
 & &
\label{215}
\end{eqnarray}

The only difference between this equation and Eq.(\ref{SD}) is the absence of the constant term $m_0 c^2$. As we indicated earlier, that constant term does not play any role in the Sommerfeld-Dirac theories and does not affect the observable energy levels. It is also interesting to note that Bohr's energy level equation was

\begin{equation}
W =  - \: \frac{m_0 e^4}{2 \hbar^2 n^2},
\label{216}
\end{equation}

and hence Eq.(\ref{215}) can be essentially regarded as an improved version of Bohr's original formula.\\
\\
\\
{\large\bf Historical Note and Acknowledgement:}\\
\\
Dr. Peter Enders of Germany has thankfully examined this paper as well as my earlier paper and remarked that $H=mv^2$ was Leibniz's ``{\em vis viva}'', or the total energy according to Leibniz. However, it should be pointed out here that Leibniz's {\em vis viva} was meant to be an expression for the kinetic energy of a body, and was subsequently modified later to become $\frac{1}{2} mv^2$ (as the kinetic energy is known today). The total energy expression $H=mv^2$ used in the present work -though it matches Leibniz's old formula- emerged from the reconciliation of two of the most important theories of the 20th century, namely, wave mechanics and special relativity, and thus it is not connected with that historical development of the 18th century.

\end{document}